\documentstyle[12pt,aaspp4]{article}
 
\input{epsf}

\journalid{}{}
\articleid{}{}
\begin{document}
\def\lax    {\ifmmode{_<\atop^{\sim}}\else{${_<\atop^{\sim}}$}\fi}
\def\gax    {\ifmmode{_>\atop^{\sim}}\else{${_>\atop^{\sim}}$}\fi}
\def\gtorder{\mathrel{\raise.3ex\hbox{$>$}\mkern-14mu
             \lower0.6ex\hbox{$\sim$}}}
\def\ltorder{\mathrel{\raise.3ex\hbox{$<$}\mkern-14mu
             \lower0.6ex\hbox{$\sim$}}}
 
\long\def\***#1{{\sc #1}}
 
\title{On the two types of steady hard X-ray states of GRS 1915+105.}

\author{Sergey P. Trudolyubov\altaffilmark{1,2}}

\altaffiltext{1}{NIS-2, Los Alamos National Laboratory, Los Alamos, NM 87545; 
tsp@nis.lanl.gov}

\altaffiltext{2}{Space Research Institute, Russian Academy of Sciences, 
Profsoyuznaya 84/32, Moscow, 117810 Russia}

\begin{abstract}
Using the data of 5 years of {\em RXTE} observations we investigate the 
X-ray spectral and timing properties of GRS 1915+105 during the {\em hard 
steady} states. The broad-band energy spectrum of the source during these 
periods is dominated by extended hard component with characteristic cut-off/
break at $\sim 10 - 120$ keV. The power density spectrum of the source 
rapid aperiodic variability shows dominant band-limited white noise 
component breaking at few Hz accompanied by a group of strong QPO peaks 
and in some cases an additional high frequency noise component with 
characteristic cut-off at $\sim 60 - 80$ Hz.

According to the results of our simultaneous X-ray spectral and timing 
analysis the behavior of the source during the {\em hard steady} states 
can be reduced to a couple of major distinct types. i){\em Type I states:} 
The dominant hard component of the energy spectrum has characteristic 
quasi-exponential cut-off at $50 - 120$ keV. The broad-band power density 
spectrum of the source shows significant high frequency noise component 
with a cut-off at $\sim 60 - 80$ Hz. ii){\em Type II states:} The hard 
spectral component has a break in its slope at $\sim 12 - 20$ keV. The 
high frequency part of the power density spectrum fades quickly lacking 
significant variability at frequencies higher than $\sim 30$ Hz. These 
two types of the X-ray hard states are also clearly distinguished by their 
properties in the radio band: while during the {\em type I} observations 
the source tends to be 'radio-quiet', the {\em type II} observations are 
characterized by high level of radio flux ({\em 'plateau'} radio states). 

In this work we demonstrate aforementioned differences using the data of 
12 representative {\em hard steady} state observations. We conclude that 
the difference between these two types can be probably explained in terms 
of different structure of the accretion flow in the immediate vicinity of 
the compact object due to presence of relativistic outflow of matter.     
\end{abstract} 

\section{INTRODUCTION}

The X-ray source GRS 1915+105, the first Galactic object found to generate 
superluminal jets (Mirabel $\&$ Rodriguez 1994), was discovered by the {\it 
GRANAT} observatory as a transient in 1992 (Castro-Tirado, Brandt $\&$ Lund 
1992). Long-term monitoring of GRS 1915+105 with the {\it RXTE} satellite 
(Bradt et al. 1993) has revealed a rich character of the source transient 
activity associated with different levels of its luminosity, which can be 
reduced to the sequence of varying {\em soft steady}, {\em flaring} and {\em 
hard steady} states qualitatively distinguished by their spectral and temporal 
properties (Greiner et al. 1996; Trudolyubov et al. 1999{\em a}; Muno et al. 
1999). It was demonstrated that the major spectral and temporal properties of 
the source during these periods are similar to that of the Galactic 
black hole candidates in the 'canonical' spectral states (Chen et al. 1997; 
Taam et al. 1997; Trudolyubov et al. 1999{\em a}).

For almost a half of the observational time the source demonstrates steady 
periods characterized by a relatively hard broad--band X-ray spectrum and 
X-ray luminosity $\sim (1 - 6) \times 10^{38} ergs/s$ (assuming the source 
distance of $12.5$ kpc) -- {\em hard steady} states (\cite{Muno99}; 
\cite{Belloni00}). In general, the energy spectrum of GRS 1915+105 during 
these periods can be approximated by a sum of the weak soft thermal component 
with characteristic color temperature $\sim 1$ keV and dominant extended 
hard component with turnover at high energies (Fig. \ref{spec_lls})
(\cite{Trudolyubov99.1}; \cite{Muno99}). The variation of the X-ray 
flux on the time scale of an individual observation does not exceed $\sim 20 
\%$. The power density spectrum (PDS) of the source is dominated by strong 
band-limited white noise (BLN) component accompanied by system of QPO 
peaks with centroid frequencies of several Hz (Fig. \ref{pds_lls}) (\cite
{Morgan97}; \cite{Chen97}; \cite{Trudolyubov99.1}). The pattern of correlation between X-ray and radio emission of the source is complicated: alternating 
periods of high (so-called 'plateau states') and low level 
radio emission associated with hard steady states have been reported 
(\cite{Foster96}; \cite{Fender99}; \cite{Muno99}).  

In this work we demonstrate that the source shows two distinct 
branches of the {\em hard steady} state (hereafter {\em type I} and {\em 
type II}), distinguished by the properties of its energy spectrum, high 
frequency X-ray variability and level of radio flux. We discuss the 
difference between these two types of the {\em hard steady} states in 
the framework of the two-phase model of the inner part of accretion flow, 
involving the outflow from the immediate vicinity of the compact object.

\section{OBSERVATIONS AND RESULTS}
The {\em RXTE} satellite performed regular pointed observations of GRS 
1915+105 in 1996 -- 2000 providing a good coverage for the detailed study of 
the spectral and timing properties of the source in the {\em hard steady} 
states. In the Table \ref{obslog} we listed the groups of observations 
covering {\em hard steady} states and individual representative observations 
used for the detailed analysis. 

\subsection{Spectral and timing analysis}
For the processing of the PCA and HEXTE data we used standard {\em RXTE} 
FTOOLS (version 5.01) tasks and methods recommended by {\em RXTE} Guest 
observer Facility.

\begin{figure}
\epsfxsize=16cm
\epsffile{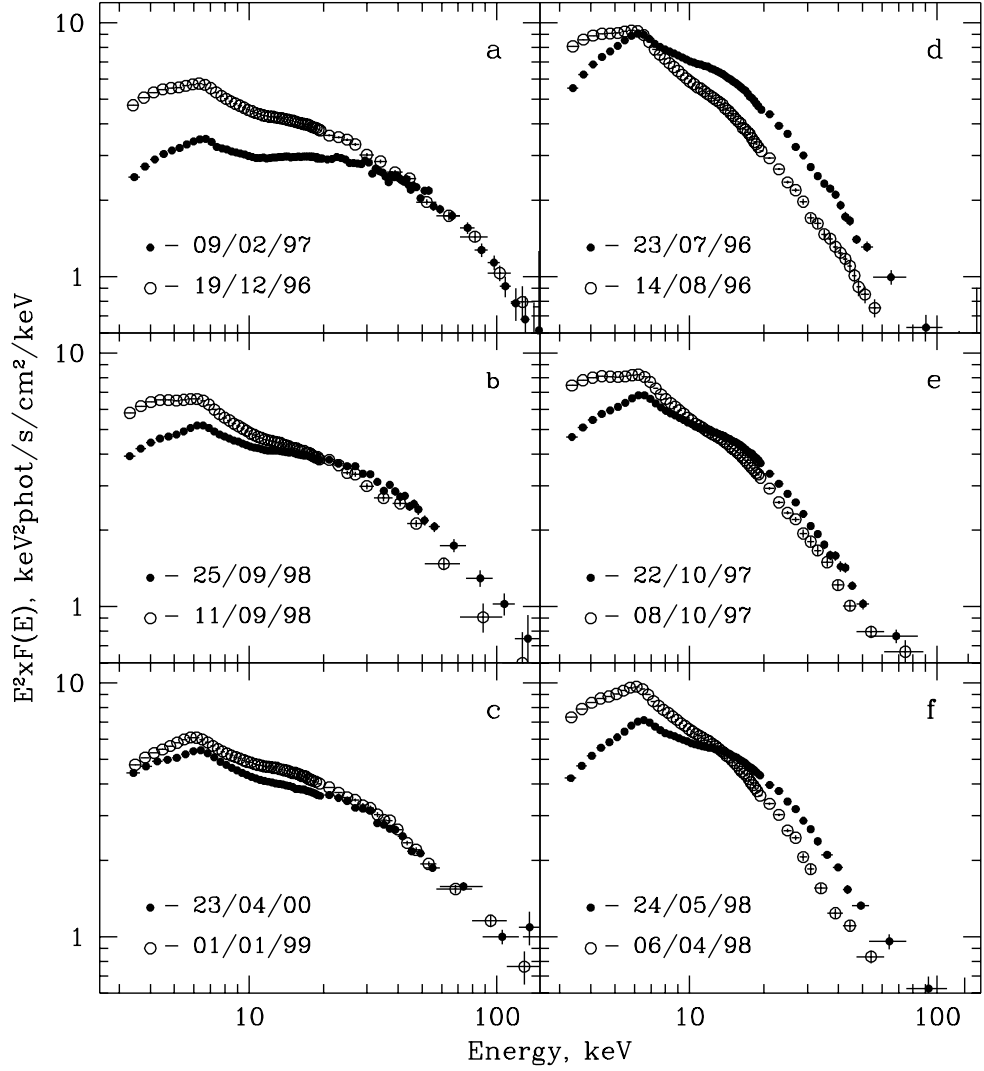}
\caption{Broad-band energy spectra of GRS 1915+105 (in units of $E^{2} 
\times F(E)$) corresponding to the {\em type I} ({\em panels a, b, c}) and 
the {\em type II} ({\em panels d, e, f}) {\em hard steady} states. \label
{spec_lls}}
\end{figure}

For spectral analysis, we used PCA {\em Standard 2} mode data, collected 
in the $3 - 20$ keV energy range. The PCA response matrices for individual 
observations were constructed using PCARMF v7.01, and background estimation 
was performed applying a Very-Large Events (VLE)--based model. Standard 
dead time correction procedure has been applied to the PCA data. Owing to the 
uncertainties of the response matrix, a 1$\%$ systematic error was added to 
the statistical error for each PCA energy channel. We used the HEXTE response 
matrices, released on October 20, 1998 and subtracted background collected in 
off-source observations for each cluster of detectors. In order to account 
for the uncertainties in the response and background determination, only 
data in the $20 - 150$ keV energy range were used for the spectral analysis. 
Typical examples of the broad-band energy spectra of GRS 1915+105 during the 
{\em hard steady} states in 1996 -- 2000 are shown in Fig. \ref{spec_lls}.

As we are interested in the general properties of the source energy spectrum 
during the {\em hard steady state}, we used only simple models to approximate 
its spectra. For the first group of observations showing a hard spectra with 
clear cut-off at the energies $\sim 60 - 90$ keV (Fig. \ref{spec_lls}, {\em 
panels a, b, c}), we used only HEXTE data in the $20 - 150$ keV energy range 
and approximated them by power law with exponential cut-off 
(Table \ref{spec_param}, {\em first part}). For the remaining observations 
(Fig. \ref{spec_lls}, {\em panels d, e, f}) data of PCA and HEXTE 
instruments covering the $3 - 150$ keV energy range were approximated by 
absorbed broken power law (Table \ref{spec_param}, {\em second part}). For 
all cases there was a distinct residuals corresponding to the iron emission
/absorption, thus Gaussian line profile and an absorption edge were included 
to refine the overall fit. An equivalent hydrogen absorbing column density 
was fixed at the level of $N_{H}L = 5 \times 10^{22}$ cm$^{-2}$.

For the timing analysis in the $2 - 30$ keV energy range the {\it RXTE}/PCA 
data in the {\em binned} and {\em event} modes containing X-ray events below 
and above 13 keV respectively were used. We generated power density spectra 
(PDS) in the $0.001 - 512$ Hz frequency range combining the results of the 
summed Fourier transforms of a short stretches of data ($8$ s) with $0.001$ s 
time bins for the $0.3 - 512$ Hz frequency range, and a single Fourier 
transform on the data in $0.125$ s time bins for lower frequencies. The 
resulting spectra were logarithmically rebinned, when necessary, to reduce 
scatter at high frequencies, and normalized to square root of fractional 
variability {\em rms}. The white noise due to the Poissonian statistics 
corrected for the dead time effects, was subtracted (\cite{zhang95}; 
\cite{Mike00}). Typical examples of the broad-band power density spectra of 
GRS 1915+105 during {\em hard steady} state (in units of $f \times 
(rms/mean)^{2}$ (\cite{Belloni97})) are shown in Fig. \ref{pds_lls}. 

Broad-band power density spectrum (PDS) of the source is dominated 
by a strong band-limited white noise (BLN) component with at least two 
characteristic breaks (at $\sim 0.1$ and $\sim$few Hz) and a complex 
of relatively narrow peaks of quasi-periodic oscillations (QPO) lying 
near the second break in the slope of BLN continuum (Fig. \ref{pds_lls}). 
For some observations (Fig. \ref{pds_lls}, {\em panels a, b, c}), 
the additional broad high frequency component with characteristic cut-off 
at $\sim 60 - 80$ Hz was also notable.

To quantify the properties of the source rapid aperiodic variability, 
PDS in the $0.05 - 256$ Hz frequency range were fitted to analytic 
models using $\chi^{2}$ minimization technique. We used a combination 
of two BLN components (approximated by zero-centered Lorentzian functions) 
and several QPO features (presented by the Lorentzian functions). For some 
observations (Fig. \ref{pds_lls}, {\em panels a, b, c}), an additional high 
frequency noise component was approximated by zero-centered Gaussian function 
with characteristic width of $\sim 60 - 80$ Hz. Parameters of the PDS 
approximation are presented in Table \ref{pds_param}.

\subsection{Two types of X-ray hard steady states}
As it is seen from Figs. \ref{spec_lls} and \ref{pds_lls}, representative 
{\em hard steady } state observations can be subdivided into two distinct 
classes with qualitatively different spectra and PDS. Further we will refer 
to these groups of observations as {\em type I} and {\em type II} states 
(see Table \ref{spec_param},\ref{pds_param}). Below we summarize 
characteristic spectral and timing properties of both groups. 

{\em Type I states.} The energy spectrum is dominated by an extended power 
law hard component with photon index $\alpha \sim 1.8 - 2.3$ and quasi-
exponential cut-off at $\sim 60 - 90$ keV (Fig. \ref{spec_lls}({\em panel a, 
b, c}); Table \ref{spec_param}). Broad-band PDS shows prominent noise 
component with average integrated {\em rms} amplitude of $\sim 4 - 5 \%$ 
and characteristic cut-off frequency of $\sim 60 - 80$ Hz, detectable up to 
$\sim 150$ Hz (Fig. \ref{pds_lls}, {\em panel a, b, c}; Table 
\ref{pds_param}). An average total fractional {\em rms} amplitude of the 
rapid aperiodic variability in the $2 - 30$ keV energy range is about 
$\sim 22 \%$ (Table \ref{pds_param}).
 
{\em Type II states.} The energy spectrum exhibits clear break in power law 
slope near $\sim 13 - 16$ keV. High energy part of the spectrum has roughly 
power law shape with almost constant photon index of $\alpha \sim 3.2 - 3.3$, 
in spite of significant changes in the slope of the lower energy part (Fig. 
\ref{spec_lls}, {\em panel d, e, f}; Table \ref{spec_param}). As in case of 
{\em type I} states, the PDS of the source is dominated by BLN and QPO 
components with characteristic peaks at a few Hz. Contrary to the {\em type 
I} states, high frequency part of the PDS in the {\em type II} state fades 
quickly lacking any significant variability above $\sim 30$ Hz. For some 
observations an additional noise component with characteristic cut-off at 
$\sim 15$ Hz was marginally detected (Table \ref{pds_param}). The average 
total fractional {\em rms} amplitude of the rapid aperiodic variability in 
the $2 - 30$ keV energy range is $\sim 5 \%$ lower than during the {\em type 
I} state observations (Table \ref{pds_param}).

\subsection{Correlated properties of X-ray and radio emission}
\begin{figure}
\epsfxsize=16cm
\epsffile{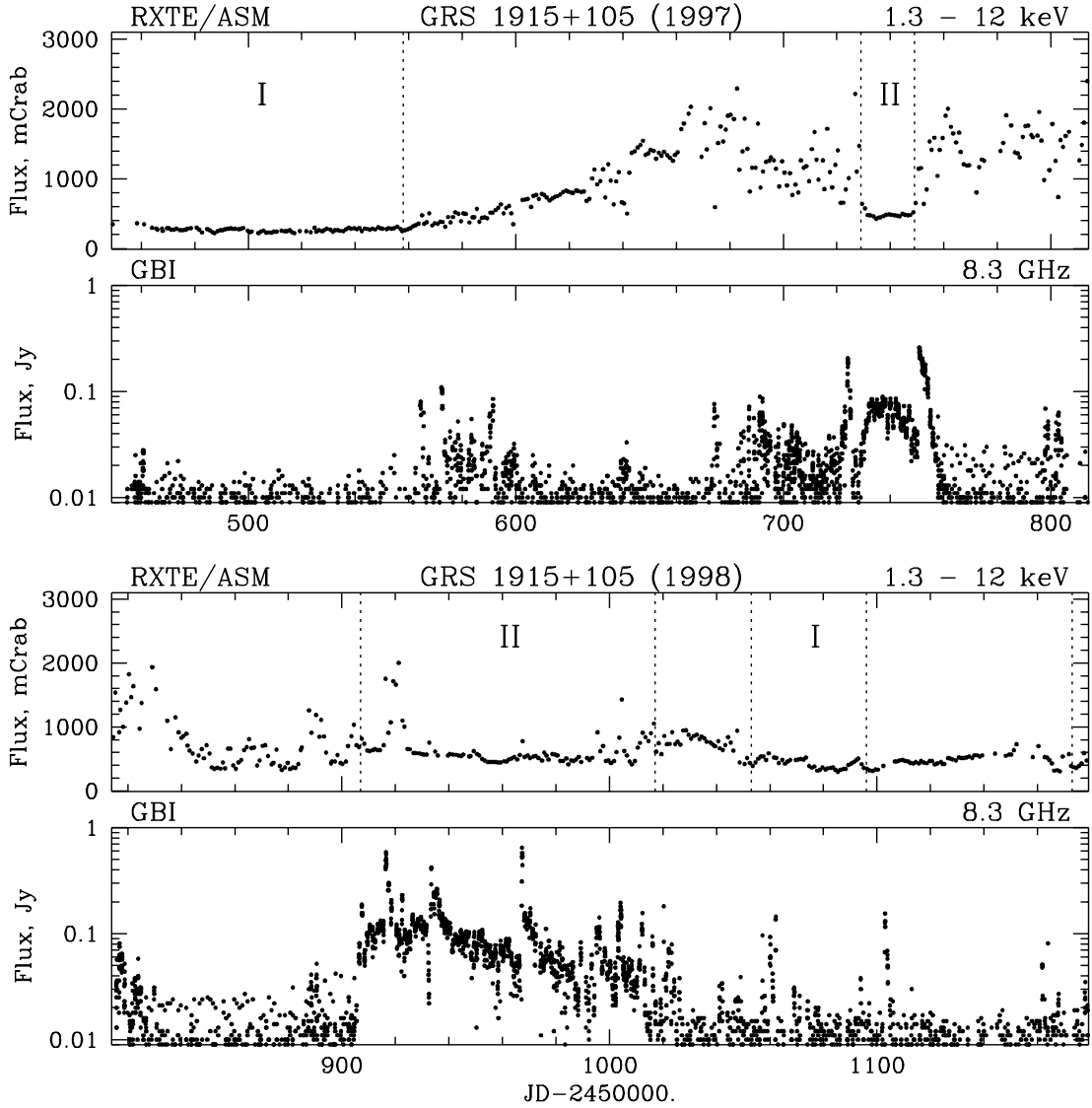}
\caption{Simultaneous X-ray and radio lightcurves of GRS 1915+105 in 1997 and 
1998. The boundaries of the {\em type I} and {\em type II} {\em hard steady} 
states are marked by dotted lines. The X-ray flux is in the mCrab units (1 mCrab $\sim$ 0.075 cnts s$^{-1}$) ({\em RXTE}/ASM and Green Bank Interferometer 
(GBI) data) \label{xray_radio_lls}}
\end{figure}

To trace the simultaneous evolution of GRS 1915+105 in radio band we 
used publicly available data of the Green Bank Interferometer monitoring 
observations at 8.3 GHz (\cite{Foster96}). Fig. \ref{xray_radio_lls} shows 
examples of the X-ray and radio lightcurves of the GRS 1915+105 
during the {\em hard steady} state in $1997 - 1998$. We note that all periods 
of {\em type I} X-ray {\em hard steady} state correspond to low level of 
radio emission (the flux at $8.3$ GHz is typically lower than $10 - 
20$ mJy). \footnote{Note, that the short 'spikes' seen on the radio 
lightcurve correspond to the X-ray {\em flaring} states, occasionally 
interrupting {\em hard steady} states} On the other hand, periods of 
{\em type II} state are characterized by much higher level of radio flux 
($\sim 50 - 100$ mJy). {\em Type II} state is also referred as a quasi-
stable {\em 'plateau'} radio state with flat radio spectrum (\cite{Foster96}; 
\cite{Fender99}) and accompanying bright infrared emission (\cite{Reba98}). 
To explain the properties of the source during {\em 'plateau'} states the 
formation of the optically thick radio source powered by outflow of matter 
from the system was proposed (\cite{Fender99}).       

\section{DISCUSSION}
To explain the observational properties of black hole binaries in the 
{\it hard} and {\it very high} states (\cite{TL95}) a number of models 
involving the hot Comptonization region near the compact object surrounded 
by the optically thick accretion disk (\cite{ss73}) were proposed (\cite
{CT95}). According to 
these models the soft thermal component of the energy spectrum is emitted 
by the optically thick accretion disk, while the hot inner region is 
responsible for generation of the hard spectral component. It was proposed 
that low and high frequency noise components in the power density spectrum 
are associated with outer optically thick and inner optically thin parts of 
the accretion flow respectively (\cite{Miyamoto94}). It is often assumed 
that QPO phenomenon is caused by the interaction between these two distinct 
parts of the accretion flow occurring on the local dynamical time scale at 
the boundary between these regions related to the local Keplerian time 
(\cite{msc96}; \cite{tlm98}). Supporting this model, the evidence of the 
correlation between an observed QPO frequency and the position of the 
boundary between optically thick and optically thin parts of the accretion 
flow was reported for the {\em flaring} states of GRS 1915+105 (\cite
{Trudolyubov99.2}; \cite{Chakrabarti00}).    

It looks natural to suggest that band limited white noise dominating the 
PDS of GRS 1915+105 may also be a product of the dynamic processes in the 
inner part of the accretion flow. Then the dynamic time scales on both inner 
and outer boundaries of this region will determine an extent of the resulting 
noise spectrum in the frequency domain. Given the position of high frequency 
cut-off in the PDS and radial dependence of the dynamic oscillation frequency, one can estimate an approximate position of the inner boundary of the 
optically thin region. Let us discuss individual properties of the two 
branches of the {\em hard steady} state on the basis of the above assumptions. We assume that characteristic frequencies of the QPO and BLN noise components 
of the power density spectra of the GRS 1915+105 are proportional to the 
Keplerian frequency at the outer and inner boundaries of the optically thin 
region of accretion flow, i.e. $f = f_{K}/ l$, where $f_{K} \approx 2200 \; 
m^{-1} r^{-3/2}$ Hz, $m$ -- mass of the compact object in solar masses, $r$ 
-- distance to the compact object in units of 3 gravitational radii.
 
\noindent {\em Type I state:} 
If the hard spectral component originates through the Comptonization of 
the low energy photons in the optically thin inner region, the observed 
cut-off energy of the spectrum indicates high electron temperature in this 
part of the accretion flow, $kT_{e} \sim 20 - 40$ keV (\cite{st80}). Typical 
value of the QPO centroid frequency allows one to estimate an approximate 
position of the outer boundary of the optically thin part of the accretion 
flow: $r_{out} \; \sim \; 170 (m f_{QPO} l)^{-2/3}$. Broad high-frequency 
component detected in all {\em type I} state observations is of particular 
interest. The position of the characteristic break of this component 
$f_{cut}^{high} \sim 60 - 80$ Hz (Table \ref{pds_param}) is remarkably close 
to the position of the reccurent stable QPO at 67 Hz observed mainly during 
the {\em soft steady} state (\cite{Morgan97}; \cite{RM98}). This fact provides 
a possibility that the broad noise component can be related to this steady 
QPO and be interpreted as the effective broadening of the coherent 
oscillation feature. Most of the models explaining the stable 67 Hz QPO 
invoke some kind of oscillations near the inner edge of the accretion 
disk in the immediate vicinity of the compact object (\cite{Morgan97} and 
references therein). These oscillations may explain the high-frequency noise 
component detected during the {\em type I} state observations.  

\begin{figure}
\epsfxsize=16cm
\epsffile{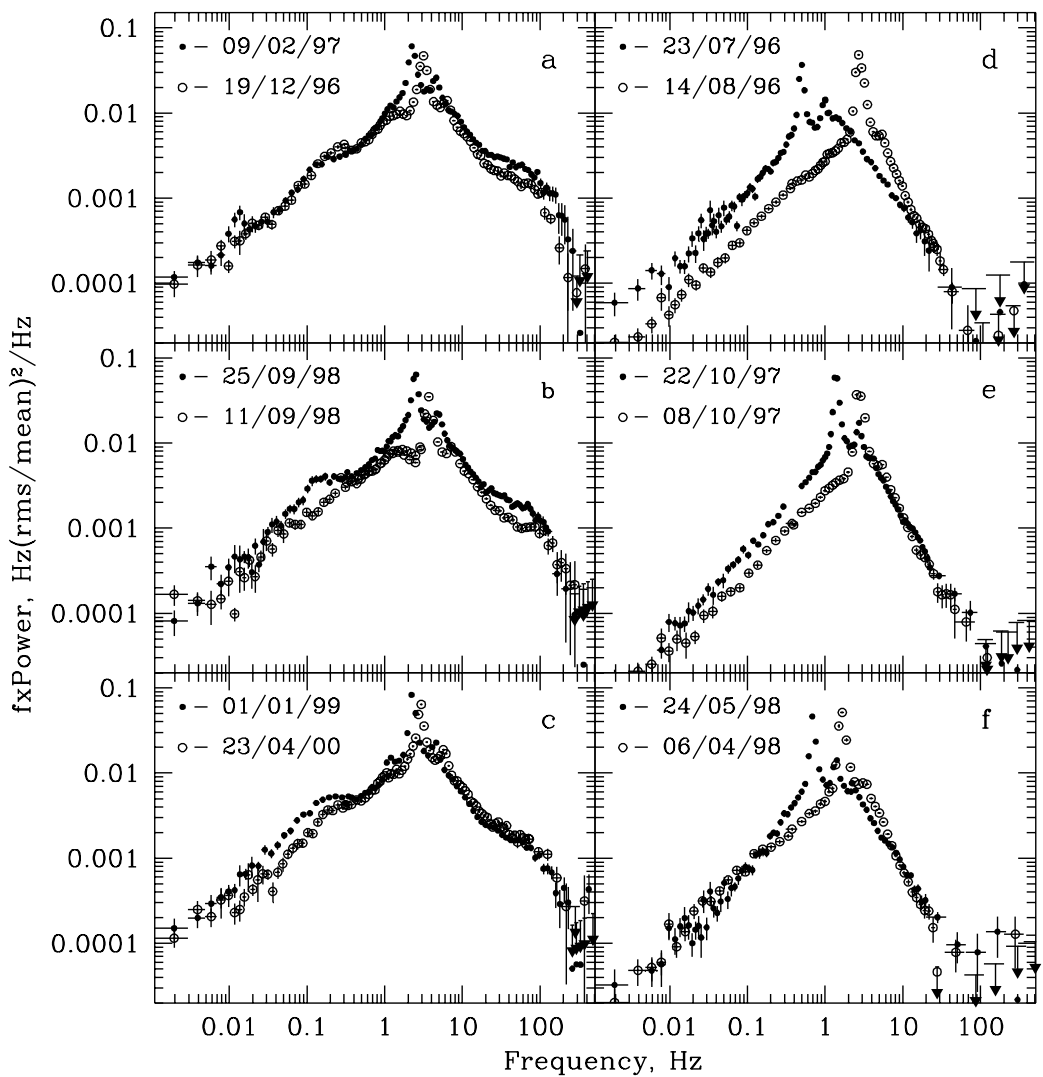}
\caption{Power density spectra (PDS) of GRS 1915+105 (in units of $f 
\times P(f)$) corresponding to the {\em type I} ({\em panels a, b, c}) and 
the {\em type II} ({\em panels d, e, f}) {\em hard steady} states. Note the 
prominent high-frequency band-limited white noise component in the {\em type 
I} PDS. \label{pds_lls}} 
\end{figure}

\noindent {\em Type II state:}
The properties of the X-ray and radio emission of GRS 1915+105 
in {\em type II} state imply significantly different geometry and 
characteristics of the accretion flow compared to that of {\em type I} 
state. We suggest two possible interpretations of the observed shape of 
the hard spectral component in the {\em type II} state. Thermal 
Comptonization mechanism implies relatively low electron temperature in 
the inner region, $kT_{e} \sim 5$ keV. On the other hand, the observed 
break in the hard component can be explained as due to the transmission 
of the hard X-rays through dense surrounding matter with average Thomson 
optical depth of the order of several (assuming power law shape of the 
spectrum of illuminating radiation) (\cite{st80}). Typical value of 
the QPO centroid frequency during the {\em type II} states is close to that 
of the {\em type I} states, indicating possible similarity in the position 
of the outer boundary of the optically thin part of the accretion flow. As 
it is seen from Fig. \ref{pds_lls}, the PDS of the {\em type II} state 
demonstrates the lack of significant variability of the source at frequencies 
above $\sim 30$ Hz. This fact may hint at the larger inner radius of the 
hot comptonization region with respect to the {\em type I} state. Assuming 
noise generation mechanism described above, and given the characteristic 
cut-off frequencies of the PDS in the {\em type I} ($f_{I}^{high} \sim 70$ Hz) and {\em type II} ($f_{II}^{high} \sim 15$ Hz) state (Table \ref{pds_param}), 
the ratio of the inner radii of the accretion flow, $r_{I}^{in}$ and $r_{II}^
{in}$ can be estimated as $(r_{II}^{in}/r_{I}^{in}) \sim (f_{I}^{high}/f_{II}^
{high})^{2/3} \sim 3$. To explain higher value of the inner radius in the 
{\em type II} state, the effective mechanism of the truncation of the 
accretion flow near the compact object in this state is needed. The 
observations at the radio and infrared wavelengths indicate the presence of 
a strong outflow of matter near the compact object coupled with a hot inner 
region (\cite{Fender99}; \cite{Eik98}). It is usually assumed that such 
outflows originate very close to the compact object, modifying the structure 
of the innermost part of the accretion flow. Finally, we may suppose that 
contrary to the {\em type I} state in the {\em type II} state the accretion 
flow extends down to some transition radius, followed by outflow region in 
the immediate vicinity of the compact object.

As a result, the difference between the properties of X-ray and radio 
emission of GRS 1915+105 during {\em type I} and {\em type II} 
{\em hard steady} states may be explained in terms of the different structure 
of the inner part of the accretion flow. Contrary to the {\em type I} X-ray 
state, in the {\em type II} state an inner region of the accretion flow 
coupled with a strong outflow of matter near the compact object, which causes 
the aforementioned differences in the observational properties.    

Finally, our two types of the {\em hard steady} states may be compared with 
the previous systems of 'states' used to characterize X-ray variability of 
GRS 1915+105. According to the definition, the 'C' state of Belloni et al. 
(2000) (with four classes of variability $\chi_{1}$, $\chi_{2}$, $\chi_{3}$, 
$\chi_{4}$) corresponds to our {\em hard steady} state. It should be 
also noted that similar distinction for a limited sample of $1996 - 
1997$ 'hard steady' observations based on the level of radio flux was 
proposed by Muno et al. (1999). The 'radio-loud hard steady state' and 
'radio-quiet hard steady state' of Muno et al. (1999) are directly 
associated with our {\em type I} and {\em type II} {\em hard steady} states. 

This research has made use of data obtained through the High Energy 
Astrophysics Science Archive Research Center Online Service, provided 
by the NASA/Goddard Space Flight Center. The Green Bank Interferometer 
is a facility of the National Science Foundation operated by the NRAO 
in support of NASA High Energy Astrophysics programs.

\clearpage

\begin{table}
\small
\caption{The list of {\it RXTE} observations of GRS 1915+105 in the {\em hard 
steady} states. The information on the individual representative observations 
is shown in column 3 (Obs. ID/Date, UT). \label{obslog}} 
\begin{tabular}{cccc}
\hline
\hline
RXTE Proposal ID & Dates, UT                      & Obs. ID/Date, UT& Ref.\\
\hline
10258, 10408     & Jul. 23, 1996 -- Sep. 03, 1996 &10258-01-01-00/Jul. 23, 1996& $1,2,4,6$\\
                 &                                &10258-01-04-00/Aug. 14, 1996& \\   
\hline
10408, 20402     & Oct. 23, 1996 -- Apr. 25, 1997 &20402-01-07-00/Dec. 19, 1996& $3,4,5,6$\\
                 &                                &20402-01-15-00/Feb. 09, 1997& \\
\hline
20187, 20402     & Oct. 05, 1997 -- Oct. 25, 1997 &20402-01-49-00/Oct. 08, 1997& $5,6$  \\
                 &                                &20402-01-51-00/Oct. 22, 1997& \\
\hline
30402, 30703     & Apr. 04, 1998 -- Jul. 27, 1998 &30703-01-14-00/Apr. 06, 1998& $7$     \\
                 &                                &30703-01-20-00/May  24, 1998& \\
\hline
30402, 30703     & Aug. 28, 1998 -- Oct. 10, 1998 &30402-01-17-00/Sep. 11, 1998& $7$ \\
                 &                                &30703-01-35-00/Sep. 25, 1998&  \\
\hline
30703, 40703     & Dec. 26, 1998 -- Mar. 27, 1999 &40703-01-01-00/Jan. 01, 1999& $7$ \\
                 & Jun. 02, 1999 -- Jun. 07, 1999 &              & $8$ \\
                 & Dec. 01, 1999 -- Feb. 21, 2000 && $7$               \\
\hline
50405, 50703     & Mar. 08, 2000 -- May. 05, 2000 &50405-01-03-00/Apr. 23, 2000& $7$ \\
\hline
\end{tabular}
\begin{list}{}{}
\item[1] -- \cite{Chen97}
\item[2] -- \cite{Morgan97}
\item[3] -- \cite{Trudolyubov99.1}
\item[4] -- \cite{Muno99}
\item[5] -- \cite{Reig00}
\item[6] -- \cite{Belloni00}
\item[7] -- this work
\item[8] -- \cite{rao00}
\end{list}
\end{table}

\clearpage

\begin{table}
\small
\caption{Spectral parameters of GRS 1915+105 in the {\em hard steady} state. 
Parameter errors correspond to $1 \sigma$ confidence level. 
\label{spec_param}}
\small
\begin{tabular}{cccccc}
\hline
\hline
\multicolumn{6}{c}{\em Type I$^{a}$}\\
\hline
Date      & Power Law   & Cut-Off energy &$\chi^{2}$&Flux$^{b}$&\\
(UT)      &   Index     &      (keV)     & (d.o.f.) &          &\\ 
\hline
19/12/1996&$2.22\pm0.06$&$85\pm15$&$130.0(115)$&$2.141$ &\\
09/02/1997&$1.83\pm0.05$&$63\pm5 $&$185.5(154)$&$1.521$ &\\
11/09/1998&$2.20\pm0.17$&$66\pm11$&$35.6(37)$  &$2.274$ &\\
25/09/1998&$2.04\pm0.07$&$66\pm7$ &$75.8(76)$  &$2.018$ &\\
01/01/1999&$2.17\pm0.06$&$66\pm8$ &$176.2(154)$&$2.215$ &\\
23/04/2000&$2.23\pm0.08$&$83\pm14$&$78.7(76)$  &$1.985$ &\\
\hline
\hline
\multicolumn{6}{c}{\em Type II$^{c}$}\\
\hline
Date      &Power Law&Break Energy&Power Law&$\chi^{2}$&Flux$^{b}$\\
(UT)      &  Index  &   (keV)    &  Index  & (d.o.f.) &          \\
\hline
23/07/1996&$2.37\pm0.01$&$14.0\pm0.3$&$3.29\pm0.01$&$237.3(193)$&$2.640$\\
14/08/1996&$2.75\pm0.01$&$13.5\pm0.5$&$3.33\pm0.02$&$274.4(233)$&$2.487$\\
08/10/1997&$2.76\pm0.01$&$15.9\pm0.5$&$3.31\pm0.02$&$161.8(116)$&$2.273$\\
22/10/1997&$2.48\pm0.01$&$16.1\pm0.1$&$3.21\pm0.01$&$382.7(233)$&$2.053$\\
06/04/1998&$2.54\pm0.01$&$13.1\pm0.3$&$3.33\pm0.02$&$274.8(153)$&$2.434$\\
24/05/1998&$2.38\pm0.01$&$15.9\pm0.1$&$3.23\pm0.02$&$130.7(116)$&$2.190$\\
\hline
\end{tabular}
\par
\begin{list}{}{}
\item[$^{a}$] -- HEXTE data, $20 - 150$ keV energy range, power law model with exponential cut-off \\
\item[$^{b}$] -- total energy flux in the $3 - 150$ keV energy range in units of $\times 10^{-8}$ erg s$^{-1}$ cm$^{-2}$\\
\item[$^{c}$] -- PCA and HEXTE data, $3 - 150$ keV energy range, absorbed broken power law model\\
\end{list}
\end{table}

\clearpage

\begin{table}
\small
\caption{The characteristics of the power density spectra of GRS 1915+105 
during the {\em type I} and {\em type II} {\em hard steady} states. $rms_
{tot}$, $rms_{QPO}$ and $rms_{high}$ represent total {\em rms} amplitude, 
{\em rms} amplitude of the fundamental QPO harmonic and  {\em rms} amplitude 
of the additional high frequency noise component integrated over the $0.05 - 
512$ Hz frequency range. $f_{QPO}$ and $f^{cut}_{high}$ denote the centroid 
frequency of the fundamental QPO peak and characteristic break frequency 
of the high frequency component. Parameter errors correspond to a $1 \sigma$ 
confidence level. 
\label{pds_param}}
\small
\begin{tabular}{cccccc}
\hline
\hline
Date &$rms_{tot}$&$f_{QPO}$&$rms_{QPO}$&$f^{cut}_{high}$&$rms_{high}$\\
 (UT) & ($\%$) & (Hz) & ($\%$) & (Hz) & ($\%$) \\
\hline
\multicolumn{6}{c}{\em Type I}\\
\hline
19/12/1996&$22.31\pm0.05$&$3.13\pm0.01$&$12.10\pm0.14$&$63\pm4$&$4.42\pm0.32$\\
09/02/1997&$24.61\pm0.07$&$2.25\pm0.01$&$12.60\pm0.19$&$69\pm3$&$5.39\pm0.32$\\
11/09/1998&$19.66\pm0.06$&$3.67\pm0.01$&$11.22\pm0.14$&$79\pm9$&$3.53\pm0.44$\\
25/09/1998&$24.84\pm0.08$&$2.45\pm0.01$&$12.86\pm0.23$&$62\pm7$&$4.86\pm0.43$\\
01/01/1999&$25.82\pm0.05$&$2.27\pm0.01$&$15.10\pm0.18$&$63\pm4$&$4.30\pm0.25$\\
23/04/2000&$24.25\pm0.08$&$2.88\pm0.01$&$12.83\pm0.20$&$80\pm9$&$4.15\pm0.51$\\
\hline
\multicolumn{6}{c}{\em Type II}\\
\hline
23/07/1996&$16.30\pm0.10$&$0.49\pm0.01$&$11.12\pm0.21$&$ ...     $&$...$\\
14/08/1996&$15.73\pm0.04$&$2.68\pm0.01$&$8.79\pm0.19$ &$16\pm2 $&$2.14\pm0.63$\\
08/10/1997&$13.81\pm0.05$&$2.62 - 2.96$&$10.71\pm0.20$&$ ...     $&$...$\\
22/10/1997&$18.16\pm0.05$&$1.39\pm0.01$&$11.45\pm0.18$&$15\pm2 $&$2.85\pm0.72$\\
06/04/1998&$16.11\pm0.08$&$1.53 - 1.69$&$12.15\pm0.31$&$ ...     $&$...$\\
24/05/1998&$16.07\pm0.09$&$0.70\pm0.01$&$10.02\pm0.25$&$ ...     $&$...$\\
\hline
\end{tabular}
\end{table}

\clearpage

\end{document}